\documentclass[aps,pre,reprint, amsmath, amssymb,superscriptaddress]{revtex4-1}

\usepackage{morefloats}
\usepackage{bm}
\newcommand{\beq}{\begin{equation}}
\newcommand{\eeq}{\end{equation}}
\newcommand{\kB}{k_{\mbox{\tiny B}}}
\usepackage[retainorgcmds]{IEEEtrantools}
\usepackage{graphicx,tikz,placeins}
\usepackage{mathrsfs}
\usepackage{amsmath,amssymb,amsfonts,physics}
\usepackage{color}
\usepackage{float}
\usepackage{times,txfonts}
\usepackage{nicefrac}
\usepackage[colorlinks=true,linkcolor=blue,urlcolor=blue,citecolor=blue,pdfusetitle]{hyperref}
\usepackage{physics}
\usepackage{soul}

\begin{document}



\title{Thermodynamics and efficiency of sequentially collisional Brownian particles: The role of drivings}

\author{Fernando S. Filho}
\affiliation{Universidade de São Paulo,
Instituto de Física,
Rua do Matão, 1371, 05508-090
São Paulo, SP, Brasil}
\author{Bruno A. N. Akasaki}
\affiliation{Universidade de São Paulo,
Instituto de Física,
Rua do Matão, 1371, 05508-090
São Paulo, SP, Brasil}
\author{Carlos E. F. Noa}
\affiliation{Universidade de São Paulo,
Instituto de Física,
Rua do Matão, 1371, 05508-090
São Paulo, SP, Brasil}
\author{Bart  Cleuren }
\affiliation{UHasselt, Faculty of Sciences, Theory Lab, Agoralaan, 3590 Diepenbeek, Belgium}
\author{Carlos E. Fiore}
\affiliation{Universidade de São Paulo,
Instituto de Física,
Rua do Matão, 1371, 05508-090
São Paulo, SP, Brasil}
\date{\today}

\begin{abstract}
Brownian particles placed sequentially in contact with distinct thermal reservoirs and subjected to external driving forces are promising candidates for the construction of reliable thermal engines. In this contribution, we address the role of driving forces for enhancing the machine performance.
Analytical expressions for thermodynamic quantities such as power output and efficiency are obtained for general driving schemes. A proper choice of these driving schemes substantially increases both power output and efficiency and extends the working regime.  Maximizations of power and efficiency, whether with respect to the strength of the force, driving scheme or both have been considered and exemplified for two kind of drivings: a generic power-law and a  periodically  drivings.

\end{abstract}

\maketitle

\section{Introduction}
The construction of nanoscale engines has received a great deal of attention and recent technological advances have made feasible the realization of distinct setups such as quantum-dots \cite{PhysRevE.81.041106}, colloidal particles \cite{rana2014single,martinez2016brownian, albay2021shift}, single and coupled systems \cite{PhysRevLett.109.190602,mamede2021obtaining} acting as working substance and others \cite{jun2014high}. In contrast to their macroscopic counterparts, their main features are strongly influenced by fluctuations when operating at the nanoscale, having several features described within the framework of Stochastic Thermodynamics \cite{groot,seifert2012stochastic,van2005thermodynamic,broeck2010a,peliti2021stochastic}. 

Recently a novel approach, coined collisional, has been put forward as a candidate for the projection of reliable thermal engines \cite{rosas1,rosas2,noa2020thermodynamics,noa2021efficient,PhysRevResearch.3.023194}. They consist of sequentially placing the system (a Brownian particle) in contact with distinct thermal reservoirs and subjected to external
driving forces during each stage (stroke) of the cycle. Each stage is characterised by the temperature of the connected thermal reservoir and the external driving force. The time needed to switch between the thermal baths at the end of each stage is neglected. Despite its reliability in distinct situations, such as
systems  interacting  only with a small fraction of the environment and those
presenting distinct drivings over each member of system \cite{benn1,maru,saga,parrondo},
the engine can operate rather inefficient depending on the way it is projected (temperatures, kind of driving and duration of each stroke). Hence the importance for strategies to enhance its performance \cite{PhysRevResearch.3.023194,noa2021efficient}.  Among the distinct approaches, we cite those  based on the maximization of power \cite{verley2014unlikely, schmiedl2007efficiency,  esposito2009universality, cleuren2015universality, van2005thermodynamic, esposito2010quantum, seifert2011efficiency, izumida2012efficiency, golubeva2012efficiency, holubec2014exactly, bauer2016optimal}, efficiency  \cite{proesmans2015efficiency,proesmans2016brownian,noa2021efficient},   low/finite dissipation \cite{PhysRevE.92.052125,PhysRevE.100.052101} or even
the assumption of  maximization via largest dissipation  \cite{purkayastha2022periodically}.

This paper deals with the above points but with a focus on a different direction, namely the optimization of the engine performance by fine-tuning the driving at each stroke. Such idea is illustrated in  
a collisional Brownian machine, which has been considered as a working substance in several works, as from the theoretical \cite{jones2015optical, albay2018optical, kumar2018nanoscale, paneru2020colloidal,li2019quantifying,mamede2021obtaining} and experimental point of views \cite{martinez2016brownian, krishnamurthy2016micrometre,blickle2012realization, proesmans2016brownian, quinto2014microscopic}.
The collisional description allows to derive   general (and exact) expressions for thermodynamic quantities, such as output power and efficiency, irrespective of the kind of driving \cite{noa2021efficient}. In order to exploit the consequences of a distinct driving each stroke and possible optimizations, two representative examples will be considered: periodically and  (generic) power-law  drivings. The former  consists of a simpler and feasible  way to drive Brownian particles out of equilibrium \cite{blickle2012realization,proesmans2016brownian,  barato2016cost,proesmans2019hysteretic,holubec2018cycling} and providing simultaneous maximizations of the engine \cite{mamede2021obtaining}. The latter has been considered  
not only for generalizing the machine performance beyond 
constant and linear drivings  \cite{noa2020thermodynamics,noa2021efficient}, 
but also to exploit the possibility of obtaining a gain by changing its form at each stroke. 

This paper is organized as follows: Sec. II presents the model and the main expressions for the thermodynamic quantities. Efficiency and optimisation is discussed in detail for both classes of drivings in Sec. III. Conclusions and perspectives are addressed in Sec. IV.

\section{Thermodynamics and Main expressions}

We focus on the simplest projection of an engine composed of only two strokes and returning to the initial step after one cycle. The time it takes to complete one cycle is set to $\tau$, with each stroke $\in\{1,2\}$ lasting a time $\tau/2$. During stroke $i$ the Brownian particle of mass $m$ is in contact with a thermal bath at temperature $T_i$ and subjected to the external force ${\tilde f_i(t)}$.

In order to introduce thermodynamic forces and fluxes, we start from the expression for the steady entropy production (averaged over a complete
period)  given by $\overline{\sigma}=\overline{\dot{Q}}_1/T_1 +  \overline{\dot{Q}}_2/T_2$, where $\overline{\dot{Q}}_1$ and $\overline{\dot{Q}}_2$ are the
exchanged heat in each stroke. 
By resorting to the first law of Thermodynamics $\overline{\dot{Q}}_1+\overline{\dot{Q}}_2=-(\overline{\dot{W}}_1+\overline{\dot{W}}_2)$ and expressing $T_1$ and $T_2$ in terms of the mean $T$ and the  difference $\Delta T=T_2-T_1$,  $\overline{\sigma}$ can be rewritten in the following (general) form:
\begin{equation}
  \overline{\sigma} =   \frac{4T^2}{4T^2-\Delta T^2}\left[   -\frac{1}{T}\left(\overline{\dot{W}}_1 + \overline{\dot{W}}_2\right) +  \left (\overline{\dot{Q}}_1 - \overline{\dot{Q}}_2\right) \frac{\Delta T}{2T^2}\right].
  \label{pi}
\end{equation}
Expressions for work and heat are obtained
by considering the  time evolution of the system probability distribution $P_i(v,t)$, governed by the following Fokker-Planck (FP) equation \cite{mariobook,tome2010entropy,broeck2010a}
\begin{equation}
\frac{\partial P_i}{\partial t} = - \frac{\partial J_i}{\partial v} -{\tilde f}_i(t)\frac{\partial P_i}{\partial v}.
\label{64}
\end{equation}
Here  $J_i$ denotes the probability current given by
\begin{equation}
  \quad J_i = - \gamma_i v P_i - \frac{\gamma_i k_\textrm{B} T_i}{m}\frac{\partial P_i}{\partial v},
  \label{645}
  \end{equation}
with  $\gamma_i$ the viscous coefficient per mass. Applying the usual boundary conditions in the space of velocities, in which both $P_{i}(v,t)$ and $J_{i}(v,t)$ vanish as $|v|\rightarrow \infty$, the time evolution of system energy  $U_i(t) = m\langle v_i^2\rangle/2 $ during each stroke corresponds to the  sum of two terms:
\begin{equation}
\frac{d}{dt}U_i(t) = -[\dot{W}_i(t) + \dot{Q}_i(t)]
\end{equation}
with the mean power $\dot{W}_i(t)$ and  heat $\dot{Q}_i(t)$ given by
\begin{eqnarray}
    \dot{W}_i(t) &=& - m \langle v_i\rangle(t) {\tilde f}_i(t); \label{work_mean}\\
    \dot{Q}_i(t) &=& \gamma_i\left( m\expval{v_i^2}(t)-k_\text{B} T_i\right).\label{heat_mean}
\end{eqnarray}
By averaging over a complete period, one recovers the first law of Thermodynamics as stated previously. Similarly, the time evolution of the system entropy $S_i(t)=- \kB \langle \ln P_i\rangle$ during stage $i$ can be expressed as a difference between two terms:
\begin{equation}
\frac{d}{dt}S_i=\sigma_i(t)-\Phi_i(t),
\end{equation}
where $\sigma_i(t)$ and $\Phi_i(t)$ are the entropy production rate and the entropy flux, respectively, whose expressions are given by
\begin{equation}
\sigma_i(t)=\frac{m}{\gamma_i T_i} \int_{}^{}  \frac{J_i^2}{P_i}dv ,\qquad {\rm and} \qquad
   \Phi_i(t)=\frac{\dot Q_i(t)}{T_i}.
  \label{eqf}
\end{equation}
Note that the right side of Eq. (\ref{eqf}) integrated over a complete period is equivalent with the relation for $\overline{\dot{\sigma }}$
given by Eq. (\ref{pi}).
Expressions for above thermodynamic quantities in terms of system parameters are obtained by recalling that such class of systems evolve to a nonequilibrium steady state whose probability distribution $P_{i}(v,t)$ for the $i-$th stage, 
satisfying Eq. (\ref{64}), is Gaussian,
\begin{equation}
P_{i}(v,t) = \exp{-(v - \langle v_i\rangle(t))^2/2b_i(t)}/\sqrt{2\pi b_i(t)} 
\end{equation}
in which the mean $\langle v_i\rangle(t)$  and variance $b_i(t)= \langle v_i^2\rangle(t) - \langle v_i\rangle^2 (t)$ are time dependent and determined by the following equations
\beq
    \frac{d \langle v_i\rangle(t)}{dt} = 
    -\gamma_i \langle v_i\rangle(t) + {\tilde f}_i(t),
\label{v1}
\eeq
and
\beq
    \frac{d b_i(t)}{dt}  = -2 \gamma_i b_i(t)  + \frac{2\gamma_i k_\textrm{B} T_i}{m},
    \label{b1}
\eeq 
respectively. In order to obtain explicit and general results, external forces are expressed in the following form:
\begin{equation}
{\tilde f}_i(t)= 
    \begin{cases}
       X_1 g_1(t), \quad t \in [0,\tau/2] \\
       X_2 g_2(t), \quad t \in [\tau/2,\tau],
    \end{cases}
\end{equation}
where $g_i(t)$ and $X_i$ account for the kind of driving and its strength at stage $i$, respectively. Continuity of $P_{i}(v,t)$ at times $t=\tau/2$ and $t=\tau$ implies
\begin{eqnarray}
   &\langle v_1\rangle(\tau/2)=\langle v_{2}\rangle(\tau/2)\;\;\;\; &; \;\;\;\;
   b_1(\tau/2)=b_{2}(\tau/2)\\
   &\langle v_1\rangle(0)=\langle v_{2}\rangle(\tau)\;\;\;\; &; \;\;\;\;
   b_1(0)=b_{2}(\tau)
\end{eqnarray}
From the above, we arrive at the following general expressions 
(evaluated for  $\kB=1$ and equal $\gamma_i=\gamma$):
\begin{widetext}
\begin{eqnarray}
   \langle v_1\rangle(t) &=& 
    X_1\int_{0}^{t} e^{\gamma(t'-t)}g_1(t') dt'+
    \frac{1}{e^{\gamma \tau}-1}\left\{X_1
        \int_{0}^{\tau/2}  e^{\gamma(t'-t)}g_1(t') dt'+ 
        X_2 \int_{\tau/2}^{\tau} e^{\gamma(t'-t)}g_2(t')dt'\right\},
        \label{v1}\\
\langle v_2\rangle(t) &=& 
    X_2\int_{\tau/2}^{t} e^{\gamma(t'-t)}g_2(t') dt'+  
   \frac {1}{e^{\gamma \tau}-1}\left\{
   e^{\gamma \tau} X_1\int_{0}^{\tau/2} e^{\gamma(t'-t)}g_1(t') dt'+ X_2 \int_{\tau/2}^{\tau} 
   e^{\gamma(t'-t)} g_2(t')dt'\right\},
   \label{v2}\\
  b_1(t) &=& -\frac{1}{m}\frac{(T_{1}-T_{2})}{\left(1+e^ {- \gamma \tau}\right)} e^{-2\gamma t}+\frac{T_{1}}{m} \;\;\;\;\;\;\; ; \;\;\;\;\;\;\; b_2(t) = -\frac{1}{m}\frac{(T_{2}-T_{1})}{\left(1+e^ {- \gamma \tau}\right)}e^{-2\gamma (t-\tau/2)}+\frac{T_{2}}{m}.  \label{b1}
\end{eqnarray}
Inserting the above expressions into Eqs. (\ref{work_mean}-\ref{heat_mean}) and averaging over a complete cycle we finally arrive at
  \begin{eqnarray}
     \label{he1}
  \overline{\dot{W}}_1 &=& -\frac{m}{\tau  \left(e^{\gamma  \tau }-1\right)} \left[X_1^2 \left(\left(e^{\gamma  \tau}-1\right) \int_0^{\tau/2}   g_1(t) e^{-\gamma t}\, \int_0^t g_1(t^\prime) e^{\gamma t^\prime} \, dt^\prime dt +\int_0^{\tau/2} g_1(t) e^{-\gamma t} \, dt \int_0^{\tau/2} g_1(t^\prime) e^{\gamma  t^\prime} \, dt^\prime\right) \right . \nonumber\\
  & + & X_1 X_2 \left . \int_0^{\tau/2} g_1(t) e^{-\gamma t} \, dt \int_{\tau/2}^{\tau } g_2(t^\prime) e^{\gamma  t^\prime} \, dt^\prime\right],\\
  {\overline {\dot Q}_1}&=&\frac{\gamma m}{\tau}\left[\int_{0}^{\tau/2} \langle v_1\rangle^2dt-\frac{1}{2\gamma m}\tanh(\gamma \tau/2)(T_1-T_2)\right],\label{he}
 \label{he2}
\end{eqnarray}
and
%
\begin{align}
\overline{\dot{W}}_2 &= -\frac{m}{\tau  (e^{\gamma  \tau }-1)}  \bigg[X_2^2 \bigg(\int_{\tau/2}^{\tau} g_2(t) e^{-\gamma  t} dt \int_{\tau/2}^{\tau} g_2(t^{\prime}) e^{\gamma  t^{\prime}}  dt^{\prime}+(e^{\gamma  \tau }-1) \int_{\tau/2}^{\tau} g_2(t) e^{-\gamma  t} \int_{\tau/2}^t g_2(t^{\prime}) e^{\gamma  t^\prime}  dt^{\prime}  dt \bigg)  .  \label{he3}  \nonumber \\
 & +    X_1 X_2 \bigg( \int_{\tau/2}^{\tau} g_2(t) e^{-\gamma  t} \, dt\int_{0}^{\tau/2} g_1(t^{\prime}) e^{\gamma  t^{\prime}} \, dt^{\prime}  + (e^{\gamma \tau}-1) \int_{\tau/2}^{\tau}e^{-\gamma}g_2(t) dt \int_{0}^{\tau/2} e^{\gamma t'}g_1(t)dt' \bigg)\bigg], \\
{\overline {\dot Q}_2}&=\frac{m\gamma}{\tau}\bigg[\int_{\tau/2}^{\tau} \langle v_2\rangle^2dt+\frac{1}{2\gamma m}\tanh(\gamma \tau/2)(T_1-T_2)\bigg],  \label{he4}
\end{align}
%
for first and second stages, respectively and $\overline{\sigma}$ is promptly obtained by inserting above expressions in Eq. (\ref{pi}). It is worth emphasizing that Eqs. (\ref{he1})-(\ref{he4}) are general and valid for any kind of drivings and temperatures. Close to equilibrium the entropy production (Eq. (\ref{pi}))  assumes the familiar \emph{flux times force} form $\overline{\sigma}\approx J_1f_1+J_2f_2+J_Tf_T$
with forces
\begin{equation}
f_1=X_1/T \;\; ; \;\; f_2=X_2/T \;\; ; \;\; f_T=\Delta T/T^2 \;
\end{equation}
($\Delta T=T_2-T_1$) and fluxes defined by
\begin{equation}
\overline{\dot{W}}_1=-TJ_1f_1 \;\; ; \;\; \overline{\dot{W}}_2=-TJ_2f_2 \;\; ; \;\; \overline{\dot{Q}}_1 - \overline{\dot{Q}}_2=2J_T.
\end{equation}
Up to first order in the forces these fluxes can be expressed in terms of Onsager coefficients $J_1=L_{11}f_1+L_{12}f_2$, $J_2=L_{21}f_1+L_{22}f_2$ and $J_T=L_{TT}f_T$ which results in
\begin{eqnarray}
   L_{11}&=& \frac{mT}{\tau  \left(e^{\gamma  \tau }-1\right)}  \left[\left(e^{\gamma  \tau}-1\right) \int_0^{\tau/2}g_1(t) e^{-\gamma  t}\int_0^t  g_1(t^\prime) e^{\gamma t^\prime} \, dt^\prime dt \, +\int_0^{\tau/2} g_1(t) e^{-\gamma t}dt \,  \int_0^{\tau/2} g_1(t^\prime) e^{\gamma  t^\prime} \, dt^\prime\right ],\\
L_{22}&=&\frac{mT}{\tau  \left(e^{\gamma  \tau }-1\right)}\left [\int_{\tau/2}^{\tau } g_2(t) e^{-\gamma  t} \, dt \int_{\tau/2}^{\tau } g_2(t^\prime) e^{\gamma  t^\prime} \, dt^\prime+\left(e^{\gamma  \tau }-1\right) \int_{\tau/2}^{\tau } g_2(t) e^{-\gamma  t}   \int_{\tau/2}^t g_2(t^\prime) e^{\gamma  t^\prime} \, dt^\prime dt \,\right ],\\
  L_{12}&=&\frac{mT}{\tau  \left(e^{\gamma  \tau }-1\right)} \int_0^{\tau/2} g_1(t) e^{-\gamma t} \, dt \int_{\tau/2}^{\tau } g_2(t^\prime) e^{\gamma  t^\prime} \, dt^\prime \;\;\;\;\;\; ; \;\;\;\;\;\;
  L_{21}=\frac{mT e^{\gamma  \tau } }{\tau  \left(e^{\gamma  \tau }-1\right)} \int_0^{\tau/2} g_1(t^\prime) e^{\gamma  t^\prime} \, dt^\prime \int_{\tau/2}^{\tau } g_2(t) e^{-\gamma  t} \, dt,
  \label{l21ew}\\
  L_{TT}&=&\dfrac{T^2}{2\tau }\tanh{\left(\dfrac{\gamma\tau}{2}\right)}.\label{ltt}
\end{eqnarray}
\end{widetext}
Two remarks are in order. First, to verify Onsager-Casimir symmetry for the cross coefficients $L_{12}$ and $L_{21}$ it is necessary not only to reverse the drivings but also to exchange the indices $1\leftrightarrow 2$ as argued in \cite{rosas2}. Second, there is no coupling between work fluxes and heat flux. That is, the cross coefficients $L_{T1}$, $L_{1T}$, $L_{T2}$ and $L_{2T}$ are absent. Hence this class of engines does not convert heat into work (and vice versa) and always loses its efficiency when the difference of temperatures $\Delta T$ between thermal baths is large, because heat can not be converted into output work \cite{noa2021efficient,mamede2021obtaining}. 

As we gonna to see next, for
the regime of temperatures we shall concern,
efficiency properties can be solely expressed  in terms of Onsager coefficients and their derivatives.

\section{Efficiency}
 As stated before, our aim is to adjust the drivings in order
 to optimize the system performance. More concretely, given an amount of energy injected to the system, whether in the form of input work $\overline{\dot{W}}_\text{in}\equiv \overline{\dot{W}}_\text{1}<0$ and/or heat $\overline{\dot{Q}}_\text{in}=\overline{\dot{Q}}_{1}\Theta(-\overline{\dot{Q}}_{1})+\overline{\dot{Q}}_{2}\Theta(-\overline{\dot{Q}}_{2})<0$ ($\Theta(x)$ denoting the Heaviside function), it is partially converted into  power output ${\cal P}\equiv\overline{\dot{W}}_\text{2}\ge 0$. A measure of such  conversion is characterized by the efficiency, given as the ratio between above quantities:
   \begin{equation}
  \eta=-\frac{{\cal P}}{{\overline {\dot W_{\rm in}}}+{\overline {\dot Q_{\rm in}}}}.
  \label{eff1}
\end{equation} 
 In order to obtain a first insight about the role driving, we shall split the analysis in two parts: engine
 operating at an unique temperature but the driving is changed each half stage. Next, we extend for different temperatures and distinct drivings.
 
\subsection{Overview about  Brownian work-to-work converters  and distinct maximizations routes}\label{sec4}
Work-to-work converters have been studied broadly in the context  of biological motors such as kinesin \cite{liepelt1,liepelt2} and myosin, in which chemical energy is converted into mechanical and vice-versa \cite{hooyberghs2013}. More recently, distinct  
work-to-work converters  made of Brownian engines  have been attracted considerable attention  and in this section we briefly revise their main aspects \cite{proesmans2016brownian,proesmans2017underdamped,noa2020thermodynamics,noa2021efficient,mamede2021obtaining}. 
Setting $\Delta T=0$ (hence $f_T=0$) reduces the efficiency Eq. (\ref{eff1})
to the ratio between Onsager coefficients:
 \cite{noa2020thermodynamics,noa2021efficient,mamede2021obtaining}
\begin{equation}
    \eta \equiv - \frac{{\cal P}}{\overline{\dot{W}}_\text{in}}=-\frac{L_{22}f_2^2+L_{21}f_1f_2}{L_{12}f_1f_2+L_{11}f_1^2}.
    \label{efff}
\end{equation}
For any kind of driving and period, the engine regime ${\cal P}>0$ implies
that the absolute value for the output force $f_2$  must lie in the interval $0\le |f_2|\le |f_m|$, where $f_m=-L_{21}f_1/L_{22}$. Since  power, efficiency and dissipation are not independent from
 each other,   $f_m$ can  be  related with $f_{2mS}$ (for $f_1$ and driving parameter constants), in which the  dissipation is minimal. It is straightforward to show that
 $f_{m}=f_{2mS}+(L_{12}-L_{21})f_1/(2L_{22})$. Note that  $f_m = f_{2mS}$ only when the Onsager coefficients are symmetric, $L_{12}=L_{21}$.

Optimized quantities, whether power and efficiency, can be obtained
under three distinct routes: optimization with respect to (i) the output force $f_2$ (keeping $f_1$ and a driving parameter $\delta$ fixed),  (ii) the driving parameter $\delta$ (forces $f_1$ and $f_2$ held fixed) and (iii) a simultaneous optimization with respect to both $f_2$ and  $\delta$. As discussed in Refs. \cite{proesmans2016brownian,noa2020thermodynamics,noa2021efficient}, such optimized quantities can expressed in terms of Onsager coefficients
and their derivatives.

The former case (maximization with respect to the output force)
is similar to findings from Refs.~\cite{proesmans2016brownian, noa2020thermodynamics,noa2021efficient}, in which the maximum power ${\cal P}_{MP,f_2}$  (with efficiency ${\eta}_{MP,f_2}$)
and maximum efficiency ${\eta}_{ME,f_2}$   (with power ${\cal P}_{ME,f_2}$) are obtained via optimal adjustments $f_{2MP}$ and $f_{2ME}$. By taking the derivative of ${\cal P}$ and Eq. (\ref{efff})
with respect to $f_2$, $f_{2ME}$ and $f_{2MP}$ are given by
\begin{equation}
  f_{2ME}=\frac{L_{11} }{L_{12} }\left(-1+ \sqrt{1-\frac{L_{12} L_{21}}{L_{11} L_{22}}}\right)f_1
  \quad {\rm and} \quad f_{2MP}=-\frac{1}{2}\frac{L_{21}}{L_{22}}f_1,
  \label{x2mp}
\end{equation}
and their associate efficiencies read
\begin{equation}
  \eta_{ME,f_2}=-\frac{L_{21}}{L_{12}}+\frac{2L_{11}L_{22}}{L_{12}^2}\left(1-\sqrt{1-\frac{L_{12} L_{21}}{L_{11} L_{22}}}\right),
  \label{etame}
\end{equation}
and
\begin{equation}
  \eta_{MP,f_2}=\frac{L_{21}^2}{4 L_{11}L_{22}-2L_{12}L_{21}},
\label{etamp}
\end{equation}
respectively. Analogous expressions for   the  power at maximum efficiency ${\cal P}_{ME,f_2}$
and the maximum power ${\cal P}_{MP,f_2}$ can obtained by inserting $f_{2ME}$ or  $f_{2MP}$ into the expression for ${\cal P}$. 
As stated  before, the second maximization to be considered is carried out for fixed output forces and  a given driving parameter $\delta$  is adjusted  ensuring maximum 
power $\delta_{MP}$ and/or efficiency $\delta_{ME}$, respectively. According to Ref. \cite{noa2021efficient,mamede2021obtaining}, they fulfill the following expressions \begin{equation}
  \frac{L'_{21}(\delta_{MP})}{L'_{22}(\delta_{MP})} = -\frac{f_2}{f_1},
  \label{eq:kappamP}
\end{equation}
and
\begin{equation}
  \begin{split}
  \Delta_{2212}(\delta_{ME}) &f_2^2 + \Delta_{2111}(\delta_{ME}) f_1^2 \\
  &+ \left[ \Delta_{2211}(\delta_{ME}) + \Delta_{2112}(\delta_{ME}) \right] f_1 f_2 = 0,  
  \end{split}
  \label{eq:kappameta}
\end{equation}
respectively, where  $L_{ij}'(\delta)\equiv\partial L_{ij}(\delta)/\partial \delta$ is the
 derivative of coefficient $L_{ij}$ respect
 to the driving $\delta$
 and $\Delta_{ijkl}(\delta) = L'_{ij}(\delta) L_{kl}(\delta) - L'_{kl}(\delta) L_{ij}(\delta)$. Associate maximum
  power and efficiency are given by
 \begin{equation}
  {\cal P}_{MP,\delta}=\frac{ L_{21}'(\delta_{MP})}{L_{22}'^2(\delta_{MP})}[L_{21}(\delta_{MP})L_{22}'(\delta_{MP})-L_{22}(\delta_{MP})L_{21}'(\delta_{MP})]f_1^2,
\end{equation}
and 
 \begin{equation}
  \eta_{ME,\delta}=- \frac{L_{22}(\delta_{ME}) f_2^2 + L_{21} (\delta_{ME})f_1 f_2}{L_{11}(\delta_{ME}) f_1^2 + L_{12} (\delta_{ME})f_1 f_2},
\end{equation}
 respectively, and expressions for  
 ${\cal P}_{ME,\delta}$ and $\eta_{MP,\delta}$
 are obtained under a similar way.  Although
 exact and valid
  for any  choice of the drivings $g_i(t)$ and forces $f_i$, Eqs.~(\ref{eq:kappamP}) and~(\ref{eq:kappameta}), in general, have to be solved numerically for achieving $\delta_{MP}$ and $\delta_{ME}$.
 
 In certain cases (as shall be explained next), it is possible to maximize the engine
 with respect to the output force $f_2$ and a driving parameter $\delta$ simultaneously, which corresponds to the crossing point between maximum
lines (power or efficiency)  with
respect to $f_2$ and $\delta$. More specifically, given the loccus of maxima $f_{2MP}/f_{2ME}$ ($\delta$ fixed) and $\delta_{MP}/\delta_{ME}$ ($f_2$ fixed), the global $\delta_{MP}^*/\delta_{ME}^*$ and  $f_{2MP}^*$/$f_{2ME}^*$ corresponds to their intersection. For instance, the global maximization of 
power is given by Eqs. (\ref{x2mp}) and (\ref{eq:kappamP}):
 \begin{equation}
  \frac{L'_{21}(\delta_{MP}^*)}{L'_{22}(\delta_{MP}^*)} = \frac{1}{2}\frac{L_{21}(\delta_{MP}^*)}{L_{22}(\delta_{MP}^*)} \qquad {\rm and} \qquad f_{2MP}^* = -\frac{1}{2}\frac{L_{21}(\delta_{MP}^*)}{L_{22}(\delta_{MP}^*)} f_1,
\end{equation}
whose (associate global) maximum power ${\cal P}^*$ and efficiency $\eta^*$read
 \begin{equation}
  {\cal P}^* =  \frac{1}{4} \frac{L^2_{21}(\delta_{MP}^*)}{L_{22}(\delta_{MP}^*)} f_1^2, \label{eq:powboth}
\end{equation}
and
\begin{equation}
  \eta^*=\frac{L^2_{21}(\delta_{MP}^*)}{4 L_{11}(\delta_{MP}^*)L_{22}(\delta_{MP}^*)-2 L_{21}(\delta_{MP}^*)L_{12}(\delta_{MP}^*)},
\end{equation}
 respectively. 

\subsection{Applications}
Once introduced the main expressions, we are now
in position for analyzing the role of driving in our two stage engine.
For instance, we shall consider two distinct (but exhibiting complementary features) cases:   generically periodically
drivings and  power-law drivings. As stated before, periodically drivings provided  simultaneous maximizations of an autonomous engine \cite{mamede2021obtaining}.  In this
constribution, we adress a simultaneous maximization for our collisional engine.
Conversely, power-law drivings has been considered in order to
generalize the machine performance beyond 
constant and linear drivings  as well as by exploiting the possibility of improving
the engine performance  via distinct drivings at each stage \cite{noa2020thermodynamics,noa2021efficient}.

\subsubsection{Generic periodically driving forces}\label{sec5}
\begin{figure*}
    \centering
         \includegraphics[width=0.98\textwidth]{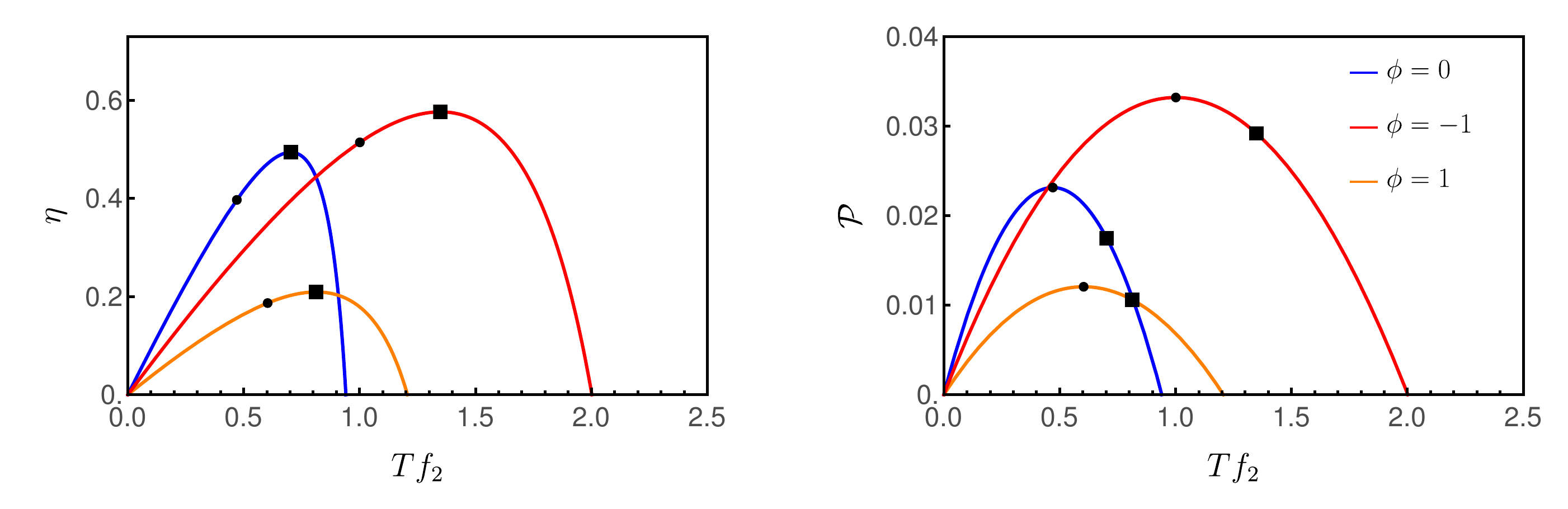} 
     \caption{For  periodic drivings, the depiction of efficiency $\eta$ and
     power-output ${\cal P}$ versus force $Tf_2$ for distinct $\phi$'s. Squares
     and circles denote the maximum efficiencies and powers, according
     to Eq. (\ref{etame})-(\ref{x2mp}). In all cases,  we set  $X_1=Tf_1=1$, $\tau=2$,  $\gamma=k_B=m=1$ and $T=1/2$.}
    \label{fig45}
\end{figure*}

\begin{figure*}
    \centering
         \includegraphics[width=0.98\textwidth]{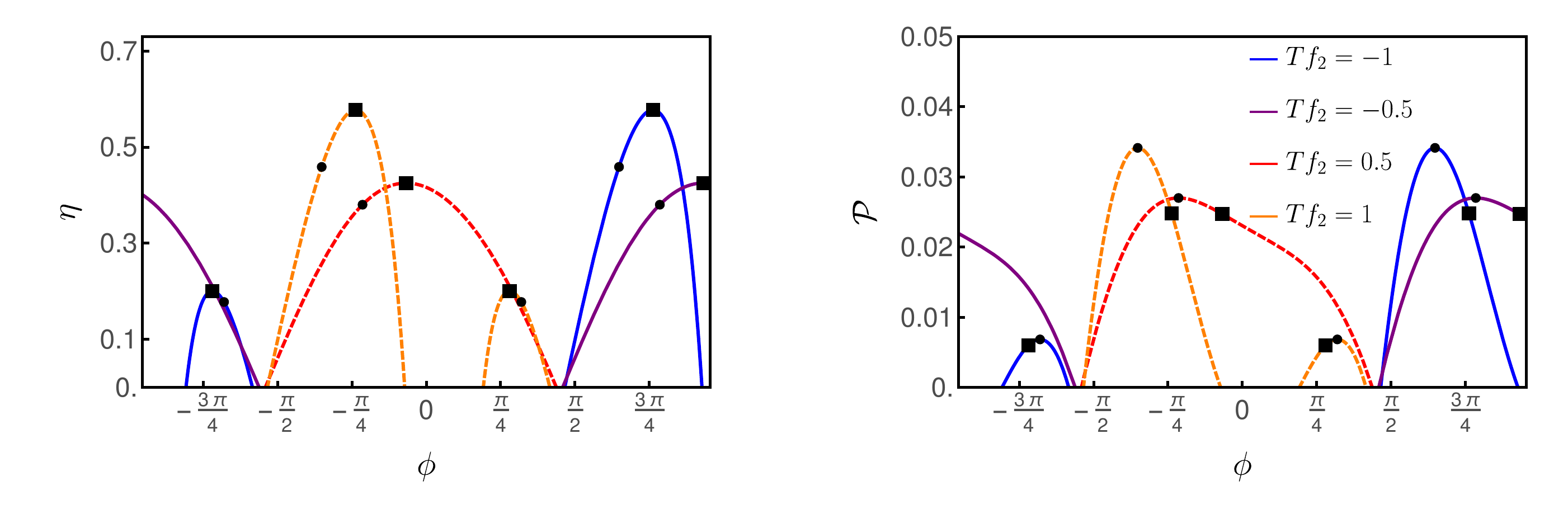} 
     \caption{For $\tau=2$ and distinct $Tf_2$'s, depiction of efficiency $\eta$
     and power-output ${\cal P}$ versus $\phi$. Squares
     and circles denote the maximum efficiencies and powers, respectively, according
     to Eqs.  (\ref{eq:kappamP}) and (\ref{eq:kappameta}). In all cases,  we set  $X_1=Tf_1=1$,   $\gamma=k_B=m=1$ and $T=1/2$.}
   \label{fig5}
\end{figure*}

\begin{figure*}
    \centering
        \includegraphics[width=0.98\textwidth]{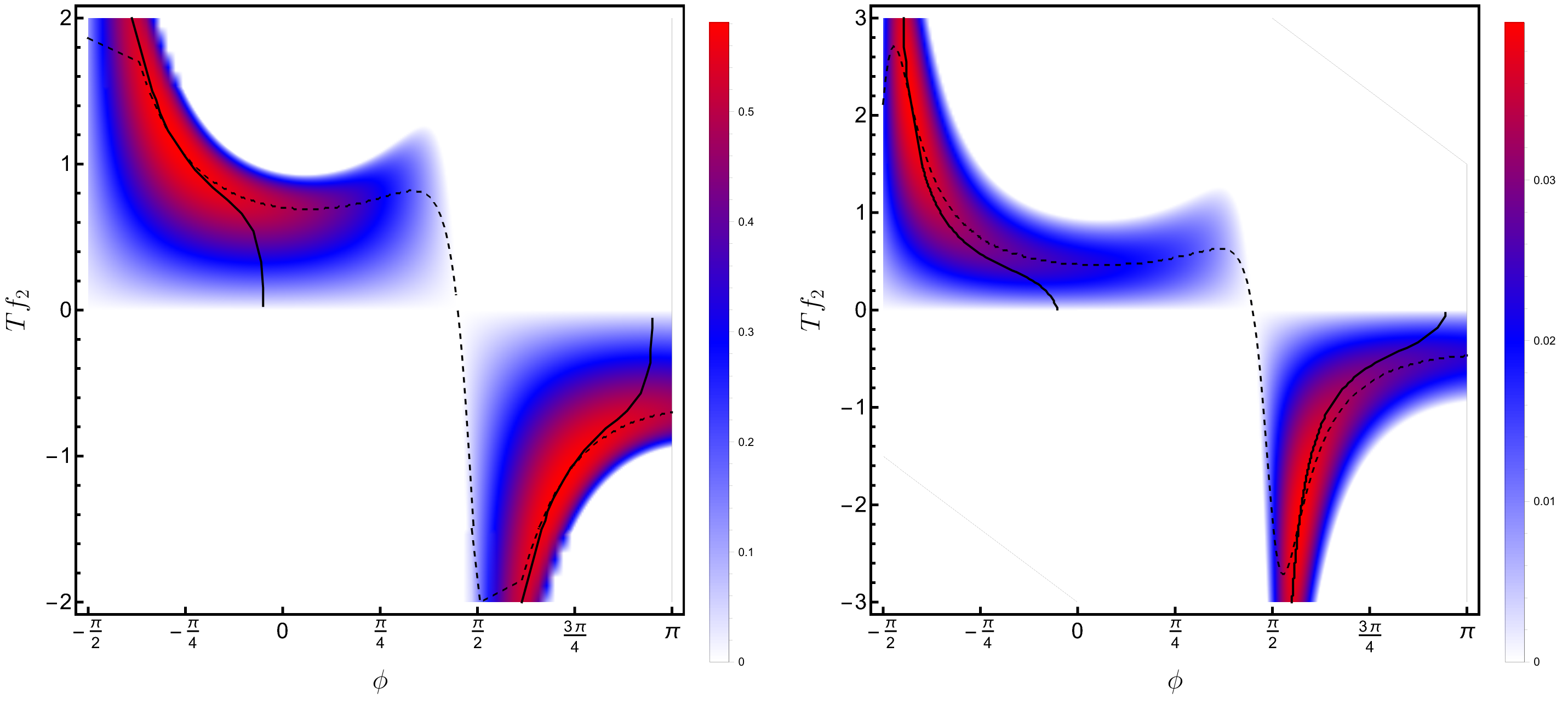}
        \caption{For the set of drivings given by Eq. (\ref{force_trigonometric}) and $\tau=2$,  left and right panels depict the phase diagram of
        the  output force $X_2=Tf_2$ versus the phase difference  
        $\phi$ for the efficiency and power output, respectively. Continuous and dashed lines denote the maximization with respect to $f_2$ and $\phi$, respectively.  In all cases,  we set  $X_1=Tf_1=1$,  $\gamma=k_B=m=1$ and $T=1/2$. }
        \label{phase}
    \end{figure*}

In this section, we apply the sequential engine under a general periodically driving, having its form and strength 
in each half stage   expressed in terms of its Fourier components:
\begin{equation}
    g_{i}(t) = 
      \sum_{n=0}^{\infty} \Big[ a_n^{(i)} \cos(\frac{4\pi n}{\tau} t) + b_n^{(i)}\sin(\frac{4\pi n}{\tau} t) \Big],
\end{equation}
for the $i-$th stage ($i=1$ or $2$),  where
 coefficients  $a_{0}^{(i)}$, $a_{n}^{(i)}$ and $b_{n}^{(i)}$ are given by
\begin{align}
    a_{0}^{(i)}&=\frac{2}{\tau} \int_{(i-1)\tau/2}^{i\tau/2}g_i(t')dt', \\
    a_{n}^{(i)}&=\frac{4}{\tau} \int_{(i-1)\tau/2}^{i\tau/2}g_i(t')\cos(\frac{4\pi n}{\tau}t')dt',\\ 
    b_{n}^{(i)}&=\frac{4}{\tau} \int_{(i-1)\tau/2}^{i\tau/2} g_i(t')\sin(\frac{4\pi n}{\tau}t')dt'.
\end{align}
Thermodynamic quantities and maximizations are also exactly obtained from Onsager coefficients,   depending 
 on Fourier coefficients $a_{0}^{(i)}$'s, $a_{n}^{(i)}$'s and $b_{n}^{(i)}$'s, whose  expressions for  generic periodically drivings are listed in 
 Appendix \ref{sin_appendix}.
 In order to tackle the role of the driving under a simpler  strategy, from now on, we shall restrict ourselves to the simplest case in which drivings at each stroke present the same frequency, but they  differ from phase difference  $\phi$ given by
 \begin{equation} 
    g_{1}(t)=
        \sin(\frac{4\pi }{\tau} t) \;\;\;\;\; ; \;\;\;\;\;
       g_{2}(t)=\sin(\frac{4\pi }{\tau} t -\phi),
    \label{force_trigonometric}
\end{equation}
The inclusion of a lag $\phi$ in the second half stage has been inspired in recent works in which it can  control of power, efficiency \cite{mamede2021obtaining} and dissipation \cite{akasaki2020entropy} and also by guiding the operation modes of the engine \cite{mamede2021obtaining}. Fig. \ref{fig45} depicts some features for distinct $Tf_2$'s and $\phi$'s, respectively.
First of all, the regime operation is delimited between 0 and $|f_{2m}|$ 
in which maximization obeys Eqs. (\ref{x2mp}) (acquiring
simpler form given by Eq. [(\ref{eqf2})] and a similar expression (although cumbersome) is obtained for $f_{2ME}/f_1$.
Note that for $\gamma\tau<<1$ and $\gamma\tau>>1$, $f_{2MP}/f_1\rightarrow 1/(2 \cos\phi)$ and $0$, respectively, such latter being independent on the lag, respectively.
\begin{widetext}
\begin{equation}
\frac{f_{2MP}}{f_1}=    \frac{4 \pi  \left(e^{\frac{\gamma  \tau }{2}}+1\right) (2 \pi  \cos (\phi )-\gamma  \tau  \sin (\phi
   ))}{\gamma  \tau  \left(\left(e^{\frac{\gamma  \tau }{2}}-1\right) \left(\gamma ^2 \tau ^2+4 \pi
   ^2\right)-4 \gamma  \tau  \left(e^{\frac{\gamma  \tau }{2}}+1\right) \sin ^2(\phi )\right)+16 \pi ^2
   \left(e^{\frac{\gamma  \tau }{2}}+1\right) \cos ^2(\phi )},
   \label{eqf2}
\end{equation}
\end{widetext}
Fig. \ref{fig5} reveals additional (and new) features coming from the lag in the second stage. The  former is
 the existence of a two distinct engine regimes for the some values of $Tf_2$,  being delimited
between two intervals $\phi_{1m}\le \phi \le \phi_{2m}$ and $\phi_{3m}\le \phi \le \phi_{4m}$ 
(fulfilling ${\cal P}=0$ at $\phi=\phi_{im}$). Also, the change of lag   moves
the engine regime  from positive to negative  of $f_2$'s. For example, for $\tau=2$ the engine regime
yields for positive (negative) output forces for $-\pi/2< \phi\le \pi/2$ ($\pi/2< \phi \le \pi$). 
 Finally, in similarity with coupled harmonic chains \cite{mamede2021obtaining}, the lag also controls the engine performance, having 
 optimal $\phi_{ME}$ and $\phi_{MP}$ in which $\eta_{ME,\phi}$ and ${\cal P}_{MP,\phi}$,
 respectively. They  obey Eqs.  (\ref{eq:kappamP})/(\ref{eqphi}) and (\ref{eq:kappameta}), the former acquire a simpler expression
\begin{equation}
\frac{f_2}{f_1}=\frac{\pi  [\gamma  \tau  \csc (\phi_{MP} )+2 \pi  \sec (\phi_{MP} )]}{\gamma ^2 \tau ^2+4 \pi ^2},
\label{eqphi}
    \end{equation}
    for the power and a more cumbersome (not shown) for $\phi_{ME}$. In the limit of $\gamma\tau<<1$   and $\gamma\tau<<1$, Eq. (\ref{eqphi}) 
  approaches to $\phi_{MP}\rightarrow \cos^{-1}(f_1/2f_2)$ and zero, 
  respectively, such latter  independent
 on the  ratio between forces, respectively.
 For completeness, Fig. \ref{phase} extends aforementioned efficiency and power findings for other values of $Tf_2$ and $\phi$. Note that suitable choices of $\phi$
and $f_2$ may lead to a substantial increase of engine performance. For example, for
$\phi=0$, the maximum $P_{MP,\delta=0}\approx 0.0231$ and $\eta_{ME,\delta=0} \approx 0.494$, whereas a simultaneous maximization 
leads to a substantial increase of power-output [given by
the intersection between Eqs. (\ref{eqf2}) and (\ref{eqphi})] $P^*\approx 0.0398$ and also
of $\eta^*\approx 0.581$.

\subsubsection{Power Law drivings}
We consider a general algebraic (power-law) driving acting at each half stage:
\beq
    g(t) =
    \begin{cases}
         \big(\frac{2 t}{\tau}\big)^{\alpha}, \textrm{ }\textrm{ for } t \in [0,\tau/2] \\
        \big(1-\frac{2t}{\tau}\big)^{\beta}, \textrm{ }\textrm{ for }t \in [\tau/2,\tau],
    \end{cases}
\eeq
where $\alpha$ and $\beta$ assume non-negative values.  
It is worth mentioning that   particular cases $\alpha=\beta=0$
and $\alpha=\beta=1$ were considered in Ref. \cite{noa2020thermodynamics}. In order to exploit in more details the influence of algebraic drivings into the first (being the worksource and heatsource) and second (responsible for the output work ${\cal P}$) stages, analysis will be carried out  by changing each  one of them separately [by kepting fixed $\beta$  and $\alpha$  in the former and latter stages, respectively]. 
Although quantities can be straightforwardly obtained from Eqs. (\ref{he1})-(\ref{l21ew}), expressions
are very cumbersome (see e.g. appendix \ref{appa}) and for this reason  analysis will be restricted for remarkable
values of $\alpha$ and $\beta$.
In principle, they can assume integer and half-integer values.
Nonetheless, inspection of exact expressions reveal that Onsager coefficients assume imaginary values when $\beta$ is half integer. Since half integer values $\alpha$ do not 
promove substantial changes (not shown), all analysis will be carried out for both $\alpha$ and $\beta$ integers.
Thermodynamics quantities are directly obtained from Eqs. (\ref{l21ew}), whose Onsager coefficients
are listed in Appendix \ref{appa}.

Fig. \ref{fig1} depicts the main portraits of 
the engine performance by varying the output force for some representative values of $Tf_2,\alpha$ and $\beta$.  Firstly, it reveals that the power output  ${\cal P}$ (left panels) is strongly (smoothly) dependent on the shape of driving acting over the first (second) half stage.
In the former case,  ${\cal P}$  is larger for smaller $\alpha$ [having its maximum for time independent ones ($\alpha=\beta=0$)] and always decreases (for all values of $Tf_2$) as $\alpha$ goes up. 
Unlike the substantial reduction of power output as $\alpha$
is raised, it is $f_2$-dependent as $\beta$ (for fixed $\alpha$) is increased, in which ${\cal P}_{mP,f_2}$ is mildly decreasing in such case. The increase of $\beta$ confers some remarkable features, such 
as the substantial increase of range of output forces (e.g. $|f_{2m}|$ increases) in which the system operates as an engine,  being restricted to positive (negative)
$f_2$'s for odd (even) values of $\beta$.
Complementary findings are achieved by examining the influence
of drivings for the efficiency. Unlike the ${\cal P}$,
$\eta$ is  $f_2$-dependent but  $\eta_{mE,f_2}$ always 
increases as $\alpha$ is raised  and  mildly decreases
as $\beta$ goes up. Finally, we stress that $\eta_{mE,f_2}$ (squares) and $\eta_{mP,f_2}$ (circles) in right panels obey Eqs. (\ref{etame}) and (\ref{etamp}), respectively, having their associate  ${\cal P}_{ME,f_2}$ and ${\cal P}_{MP,f_2}$ illustrated in the left panels.
\begin{figure*}
    \centering
    \includegraphics[width=0.8\textwidth]{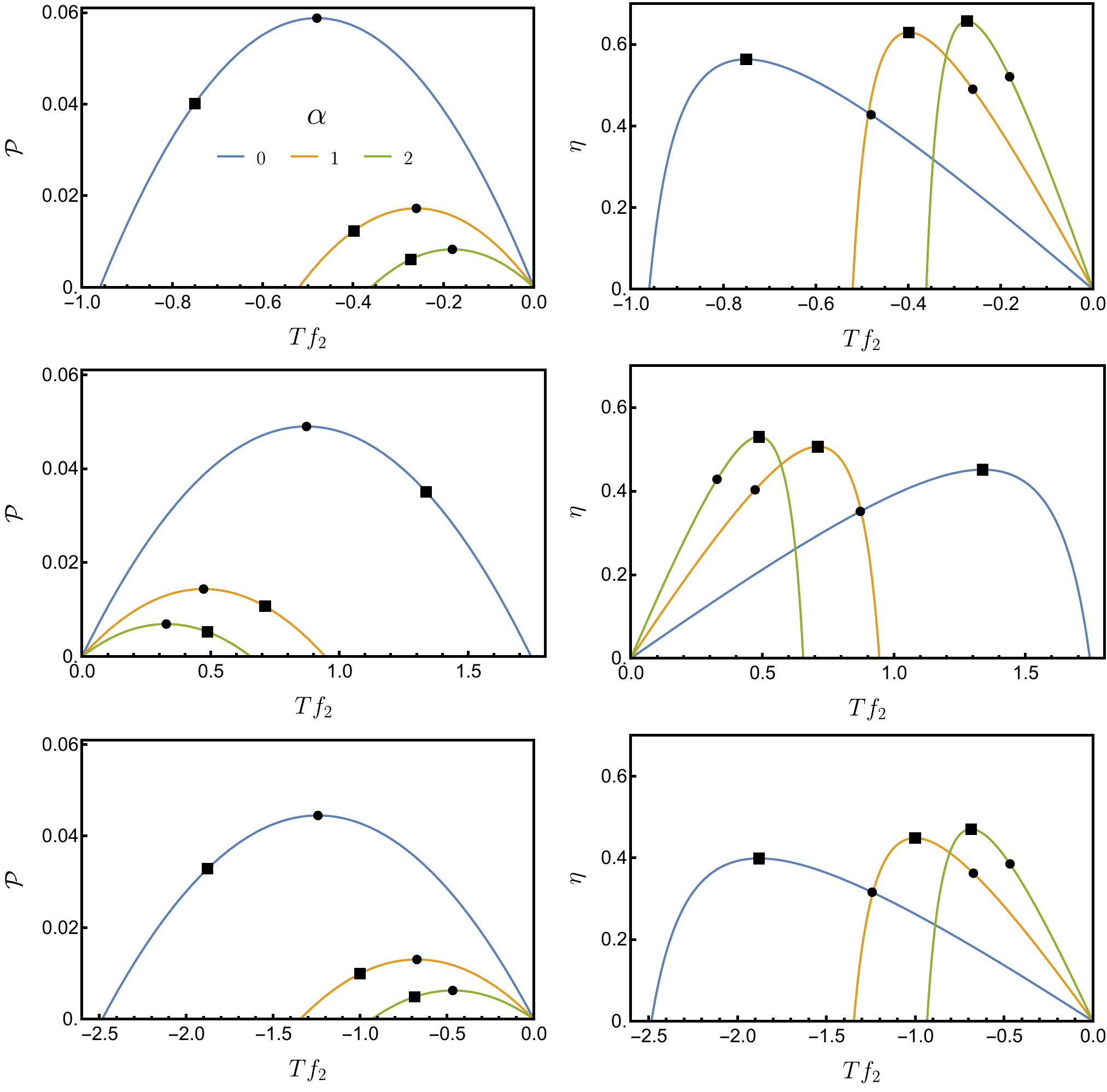} 
    \caption{For power-law drivings, the depiction of power output ${\cal P}$ and efficiency $\eta$ versus $X_2=Tf_2$  for representative values of $\alpha$  and  $\beta$ (from top to bottom, $\beta=0,1$ and $2$).  Squares
     and circles denote the maximum efficiencies and powers, according
     to Eq. (\ref{etame})-(\ref{x2mp}). 
       In all cases, we set  $X_1=Tf_1=1$, $\tau=\gamma=k_B=m=1$ and $T=1/2$.}
    \label{fig1}
\end{figure*}

\begin{figure*}
    \centering
    \begin{tabular}{cc}
         \includegraphics[width=1\textwidth]{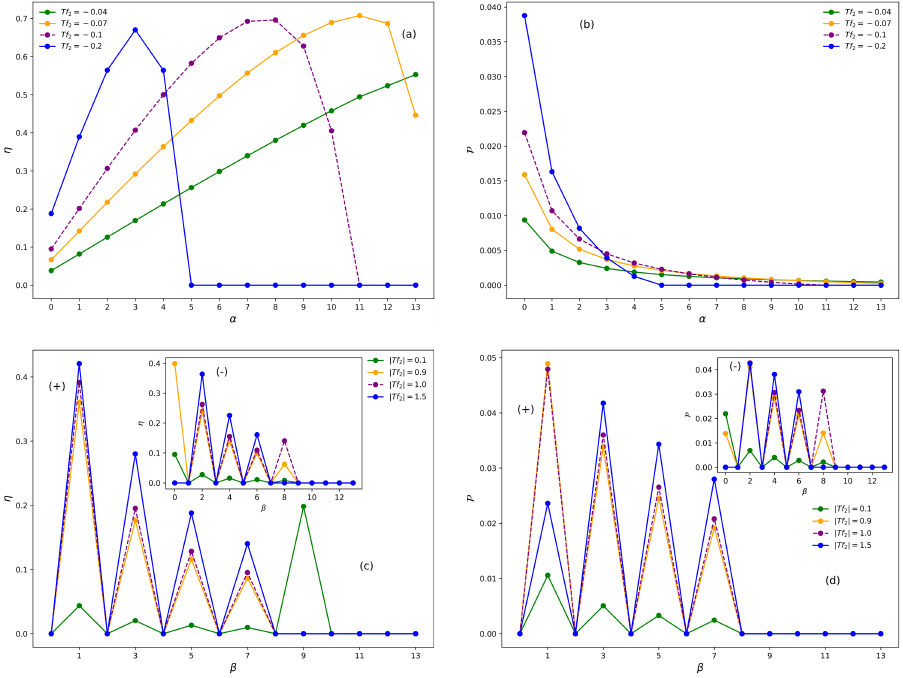}
\end{tabular}
     \caption{For fixed forces $X_2=Tf_2$, the depiction of efficiency  $\eta$ (a) and  ${\cal P}$ (c) versus  $\alpha$ ($\beta=0$). Panels (b) and (d) show the
     same, but $\beta$ is varied (for fixed $\alpha=0$).   Main panels (insets) in (c) and (d)
     show results for odd (even) $\beta$. In all cases, we set  $X_1=Tf_1=1$, $\tau=\gamma=k_B=m=1$ and $T=1/2$.}
    \label{fig2}
\end{figure*}

Next, we tackle the  opposite route, in which $f_2$ is kept fixed with $\alpha$ or $\beta$  being varied
in order to ensure optimal performance. Maximization 
of quantities follow theoretical predictions from Eqs. (\ref{eq:kappamP}) and (\ref{eq:kappameta}) and Fig. \ref{fig2}  depicts main
trends for some representative values of $f_2$. 

The dependence of driving $\alpha$ upon the efficiency shares some similarities
when compared with $f_2$ (panel $(a)$), leading to the existence of an optimal driving $\alpha>0$ ($\alpha=0$) for low (large) values of $|f_2|$. Hence, a driving beyond the constant case in the first stage can be important  for  increasing efficiency, depending on the way the machine is projected.
On the other hand, for power-output purposes, ${\cal P}_{MP,\alpha=0}$'s are always maxima  and  decreases for $\alpha>0$, irrespective to the value of output force (panel $(c)$). 

\begin{widetext}

\begin{figure}
    \centering
         \includegraphics[width=0.8\textwidth]{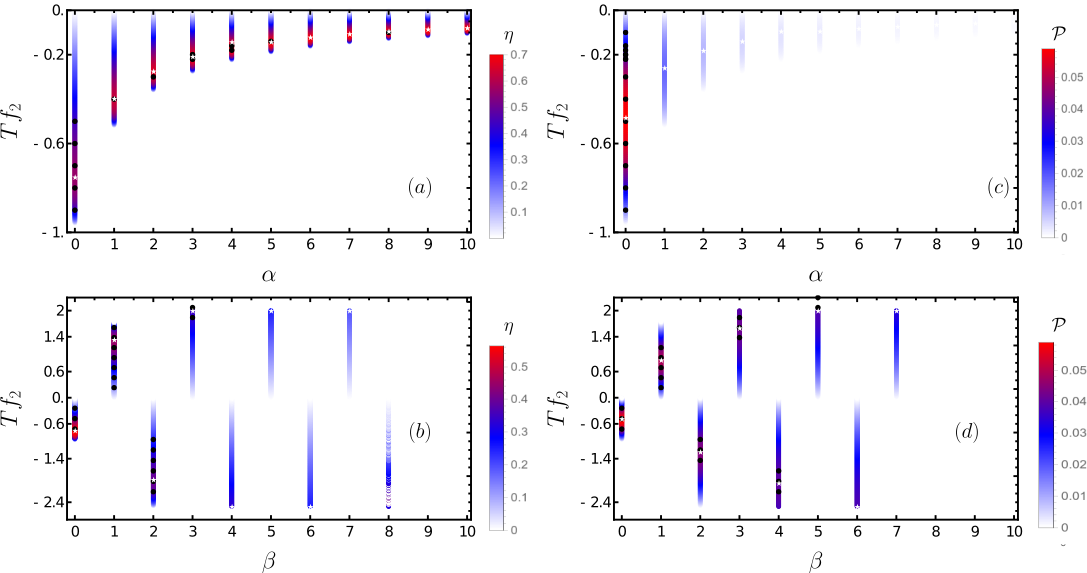} 
     \caption{Panels $(a)$ and $(c)$ depict, for $\beta=0$, the phase diagram $X_2=Tf_2$ versus $\alpha$ by considering  the efficiency (a) and power (c).
     In  (b) and (d), the opposite case ($\beta$ is varied for fixed $\alpha=0$). White
     and black symbols
     denote some representative maximizations with respect to the $f_2$ and driving (for $f_2$ held fixed), respectively. In all cases, we set  $X_1=Tf_1=1$, $\tau=\gamma=k_B=m=1$ and $T=1/2$.}
    \label{fig3}
\end{figure}
\end{widetext}

The opposite case  (fixed $\alpha$ and $\beta$ is varied)  is also more revealing and it is $f_2$ dependent (see e.g. panels 
$(c)-(d)$ and Fig. \ref{fig3}), in which
maximum efficiencies $\eta_{mE,\beta}$ and powers ${\cal P}_{mP,\beta}$ also follow Eqs. (\ref{eq:kappameta}) and (\ref{eq:kappamP}).
Fig. \ref{fig3} extends aforementioned findings for several values
of $\alpha$ and $\beta$. In contrast
 the periodically drivings, a global optimization
 in such case has not been performed, since $\alpha$ and $\beta$ present only
integer values.

 Summarizing above findings: While low $\alpha$'s 
 stage is always more advantageous for enhancing the power output, there is
 a compromise between force $|f_2|$ and $\alpha$ and $\beta$ in order to enhance the efficiency. On the other hand, although maximum efficiencies and powers
 smoothly decreases with $\beta$,  the set of output forces $|f_2|$ in which
 the system operates as an engine enlarges substantially.

\newpage

%
%

\subsection{Difference of temperatures}
In this section we examine the effect of different drivings each stroke
when the temperatures are also different. Since  ${\cal P}$ does
not depend on the temperature, the numerator of Eq. (\ref{eff1}) is
 the same as before, but now the system can receive heat
from the thermal bath 1 or 2 if $T_1>T_2$ or $T_1<T_2$, respectively and hence   such kind of engine becomes less efficiently when the difference of temperatures raises. We wish to investigate the interplay
between parameters as a strategy for compensating above point.
Taking
into account that ${\overline {\dot Q}_1}$ (or ${\overline {\dot Q}_2}$) also depends on the $f_1$ and $f_2$ [appearing inside $\langle v_1\rangle(t)$ (or $\langle v_2\rangle(t)$)], for small $\Delta T$ the system will receive heat from a thermal bath 1 (2)
only if $2m\gamma\int_0^{\tau/2}\langle v_1\rangle^2dt<\tanh(\gamma \tau/2)(T_1-T_2)$ [$2m\gamma\int_{\tau/2}^{\tau}\langle v_2\rangle^2dt<\tanh(\gamma \tau/2)(T_2-T_1)$].
Otherwise,  the thermal  engine will
behave as a work-to-work converter   and all previous analysis and expressions can be applied.
For large $\Delta T$ (not considered here), above inequalities are always fulfilled and hence  the
system efficiency is always lower than the work-to-work case.

\begin{figure*}
    \centering 
    \includegraphics[width=0.9\textwidth]{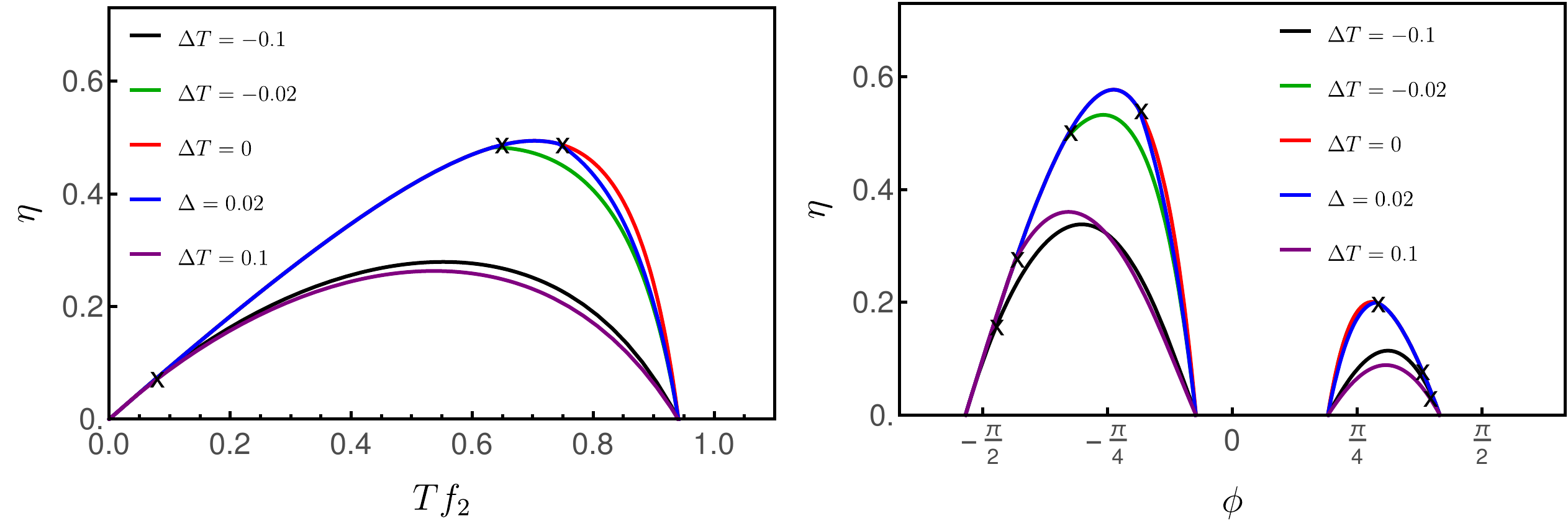} 
    \caption{ For $\tau=2,\phi=0$ and $Tf_2=1$, the depiction of 
    efficiency $\eta$ versus $Tf_2$
    and $\phi$ for  distinct temperature difference $\Delta T$ between thermal baths, respectively. Symbols $\times$ attempt to the separatrix between the work-to-work and thermal engines, respectively.
    In all cases,  we set  $T_1=1/2$, $T_2=1/2+\Delta T$, $X_1=Tf_1=1$ and $\gamma=k_B=m=1$.}
    \label{diftemp2}
\end{figure*}

\begin{figure*}
    \centering
    \includegraphics[width=1.0\textwidth]{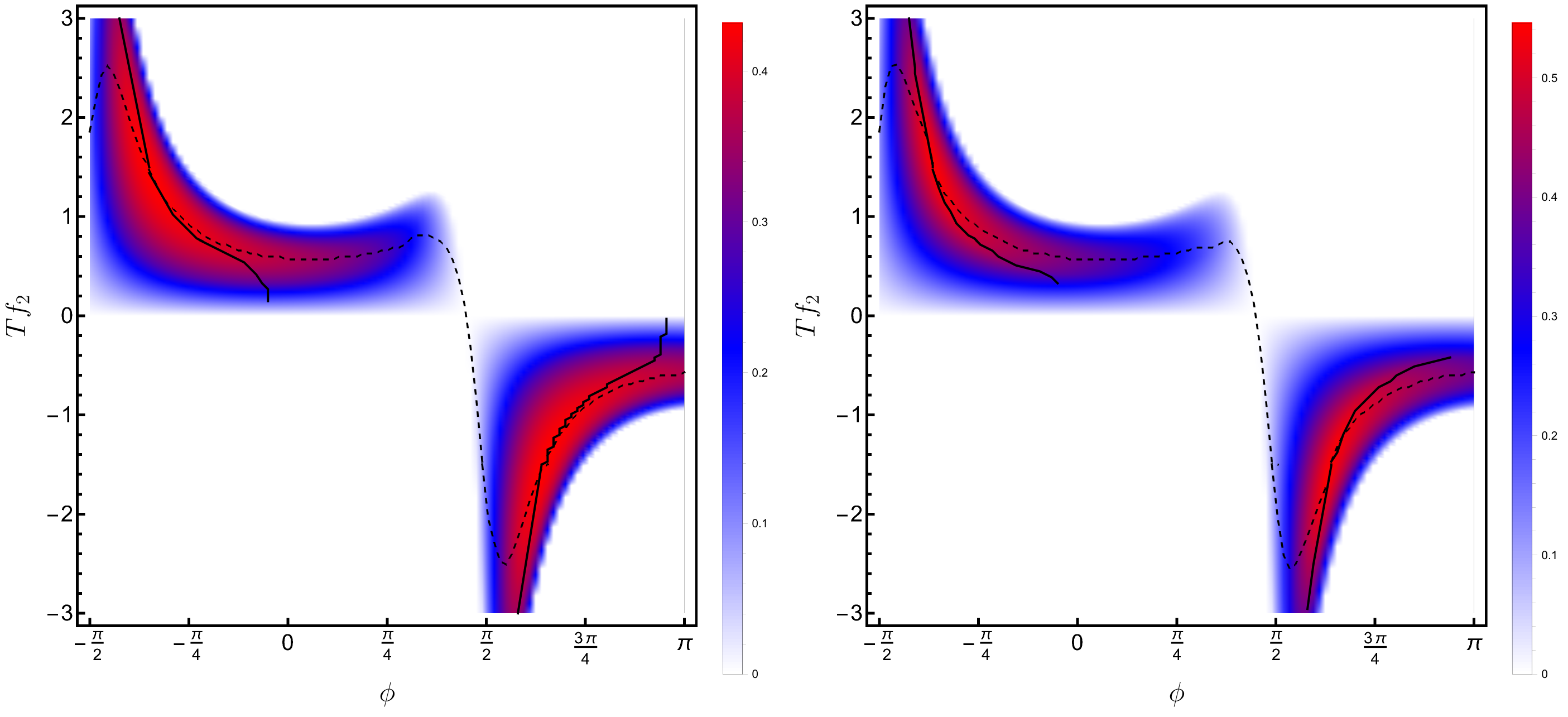} 
               \caption{For the set of drivings given by Eq. (\ref{force_trigonometric}) and $\tau=2$,  left and right panels depict efficiency phase diagrams of
        the  output force $X_2=Tf_2$ versus the phase difference  
        $\phi$ for $\Delta T=0.1$  and $-0.1$, respectively. Continuous and dashed lines denote the maximization with respect to $f_2$ and $\phi$, respectively.  In all cases,  we set  $X_1=Tf_1=1$,  $\gamma=k_B=m=1$ and $T=1/2$. }
        \label{phase2}
    \end{figure*}

Figs. \ref{diftemp2} and \ref{phase2} exemplify thermal engines for periodically drivings. As stated before, although the system operates in a similar way to 
the work-to-work converter for some values of $Tf_2$ (see e.g.  symbols $\times$
separating the thermal from work-to-work regimes), the efficiency decreases  as $\Delta T$ is raised, illustrating the no conversion of heat  into output work. 
Interestingly, the system placed in contact with the hot thermal bath in the first stage leads to somewhat higher efficiencies 
($\eta^*\approx 0.547$) than in the first stage $\eta^*\approx 0.433$). This can be understood by examining the first term in the right sides of 
Eqs. (\ref{he2}) and (\ref{he4}). Since the contribution coming from the difference of
temperatures is the same in both cases, the interplay between lag and driving forces
leads to  $\int_{0}^{\tau/2} \langle v_1\rangle^2(t)dt$ be larger than $\int_{\tau/2}^{\tau} \langle v_2\rangle^2(t)dt$ and hence confering some advantage when $T_1>T_2$.

\begin{figure*}
    \centering
    \includegraphics[width=0.8\textwidth]{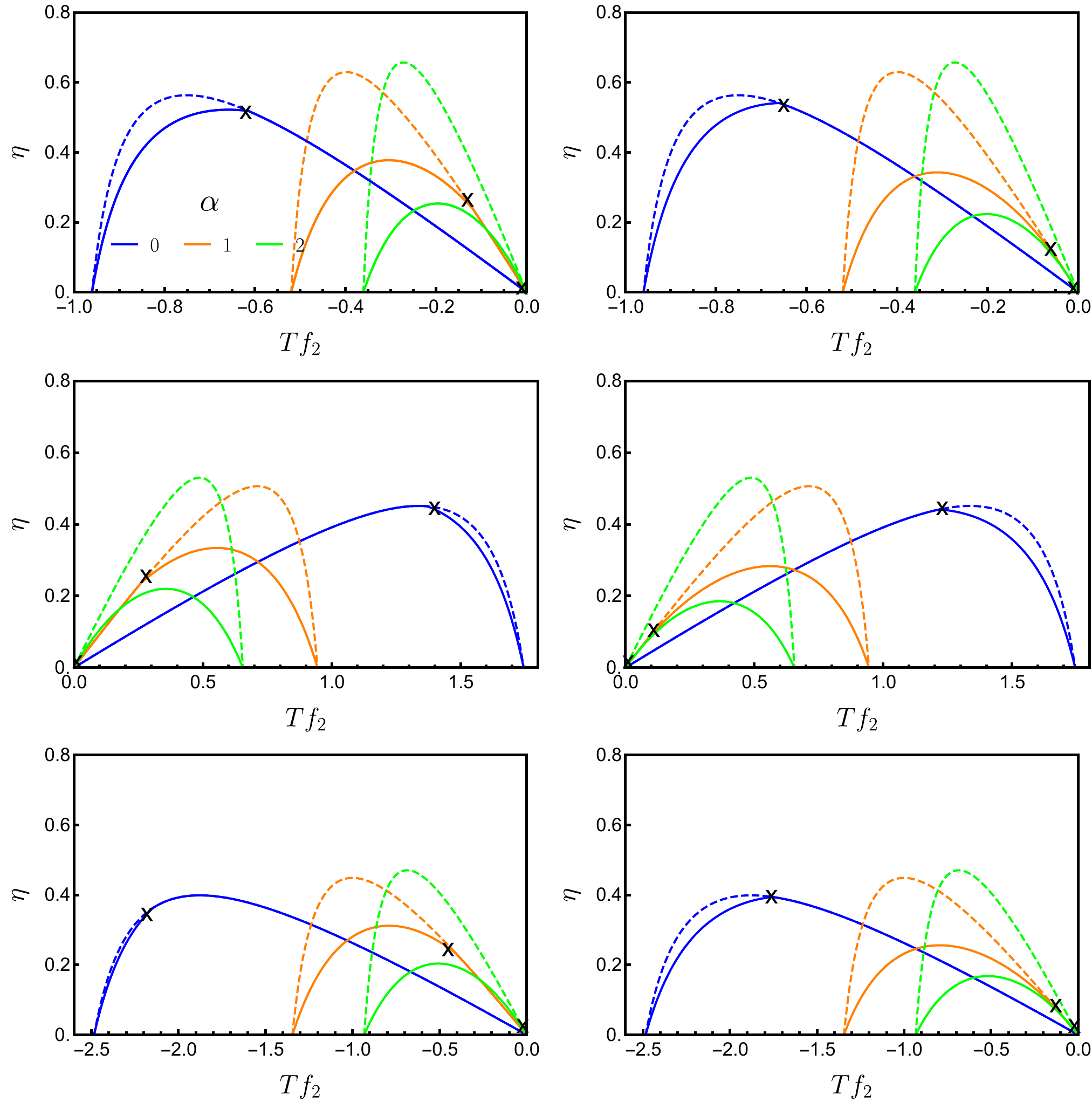} 
    \caption{Left and right panels  show  the efficiency $\eta$ versus $Tf_2$ for distinct
    $\alpha$'s for $\Delta T=0.025$ and $\Delta T=-0.025$, respectively.  Symbols $\times$ denote the separatrix between 
    the work-to-work converter (dashed lines) from the heat engine (continuous lines). From
    top to bottom, $\beta=0,1$ and $2$.}
    \label{diftemp}
\end{figure*}

Lastly, Fig. \ref{diftemp} extends the results for thermal engines
for power-law drivings.
For all values of $\alpha$ and $\beta$, the thermal engine is marked by  a reduction of its performance as $\Delta T\neq 0$, being more substantial as $\alpha$ is raised and less sensitive  to the increase of driving in the second state (increase of $\beta$).  Lastly, the engine performances also exhibit some (small) differences when the hot bath
acts over the first and second strokes (see e.g. left and right panels), being
somewhat larger when $\Delta T>0$.

  \section{Conclusions}
 The influence of the driving in a collisional approach for Brownian engine, in which the particle is subjected  each half stage  to
 a distinct force and driving, was investigated from the framework of stochastic Thermodynamics. General and exact expressions
for thermodynamic quantities, such as output power and efficiency were obtained, irrespective the kind of driving, period and temperatures. Distinct routes 
for the maximization of  power and efficiency were undertaken, whether with respect to the strength force, driving and both of them  for two kind of drivings:  generic power-law and  periodically  drivings.
The engine performance can be strongly affected when one considers simple (and different) power law drivings acting over the system at each stage. While a
 constant driving is always more advantageous for enhancing the power output, a convenient compromise between force $|f_2|$, $\alpha$ and $\beta$ can be adopted for improving the efficiency.  
Conversely, periodically drivings not only allows to perform a simultaneous
maximization of engine, in order to obtain a larger gain, but also the change of driving  (exemplified by  a phase difference in the second stage) confers a second advantage, in which the engine regime is  (substantially) enlarged the engine regime over distinct sets of parameters.


As a final comment, it is worth pointing out the decreasing of engine performance
as the difference of temperatures between thermal baths is increased. The inclusion
of new ingredients, such as a coupling between velocity and drivings, may be a candidate in order to circumvent this fact and  be responsible for a better performance of such class of collisional thermal engines.

\section{Acknowledgments}
C.E.F. acknowledges the financial support from São Paulo Research Foundation (FAPESP) under Grants No. 2021/05503-7 and  2021/03372-2. The financial support from CNPq is also acknowledged. This study was supported by the Special Research Fund (BOF) of Hasselt University under grant BOF20BL17.

\appendix

%
%
%
%
%

%
\section{Onsager coefficients for  generic periodically driving}
\label{sin_appendix}
Onsager coeficients for a generic  periodic driving in each half stage are listed below: 
\begin{widetext}
\begin{align}
    L_{11} =
    \frac{mT}{\tau (e^{\gamma \tau}-1)}
    \sum_{n=0}^{\infty}\sum_{m=0}^{\infty} &
    \bigg\{
       (e^{\gamma \tau}-1)  \int_{0}^{\tau/2} [a_m^{(1)} \cos (\frac{4\pi m}{\tau}t) + b_{m}^{(1)}\sin (\frac{4\pi m}{\tau}t)] 
        [a_n^{(1)}C_{n}^{(1)}(t) + b_{n}^{(1)}S_{n}^{(1)}(t)]
        e^{-\gamma t} dt + \nonumber\\
        &+[a_n^{(1)}C_{n}^{(1)}(\tau/2)+b_{n}^{(1)}S_{n}^{(1)}(\tau/2)]
         [a_{m}^{(1)}\overline{C}_m^{(1)}(\tau/2)+b_{m}^{(1)} \overline{S}_{m}^{(1)}(\tau/2)],
    \bigg\}
\end{align}

\begin{align}
    L_{22}=
    \frac{mT}{\tau (e^{\gamma \tau}-1)}
    \sum_{n=0}^{\infty}\sum_{m=0}^{\infty} &
    \bigg\{
       (e^{\gamma \tau}-1) \int_{\tau/2}^{\tau}
        [a_{m}^{(2)}\cos( \frac{4\pi m}{\tau}t)+b_{m}^{(2)} \sin(\frac{4\pi m}{\tau}t)]  
        \{a_n^{(2)} [C_{n}^{(1)}(t) -C_{n}^{(1)}(\tau/2)]+ b_{n}^{(2)}[S_{n}^{(1)}(t)-S_{n}^{(1)}(\tau/2)]\}
       e^{-\gamma t} dt \nonumber \\ & 
        +\{a_{n}^{(2)}[C_{n}^{(1)}(\tau)-C_{n}^{(1)}(\tau/2)]+ b_n^{(2)}[S_{n}^{(1)}(\tau)-S_{n}^{(1)}(\tau/2)]\}
        \{a_{m}^{(2)}[\overline{C}_m^{(1)}(\tau)-\overline{C}_m^{(1)}(\tau/2)]+b_{m}^{(2)}[\overline{S}_m^{(1)}(\tau)-\overline{S}_m^{(1)}(\tau/2)]\}
    \bigg\}
\end{align}

\begin{equation}
    L_{12} =
    \frac{mT}{\tau(e^{\gamma \tau}-1)}
    \sum_{n=0}^{\infty}\sum_{m=0}^{\infty} \{a_{n}^{(2)}[C_n^{(1)}(\tau)-C_n^{(1)}(\tau/2)] + b_{n}^{(2)}[S_n^{(1)}(\tau)-S_n^{(1)}(\tau/2)]\}
    [a_{m}^{(1)}\overline{C}_m^{(1)}(\tau/2)+b_{m}^{(1)}\overline{S}_{m}^{(1)}(\tau/2)]
\end{equation}

\begin{equation}
    L_{21} =
    \frac{mTe^{\gamma \tau}}{\tau (e^{\gamma \tau}-1)}
    \sum_{n=0}^{\infty}\sum_{m=0}^{\infty}
    [a_n^{(1)}C_{n}^{(1)}(\tau/2) + b_n^{(1)}S_n^{(1)}(\tau/2)]  \{a_{m}^{(2)}[\overline{C}_m^{(1)}(\tau)-\overline{C}_m^{(1)}(\tau/2)]+b_{m}^{(2)}[\overline{S}_m^{(1)}(\tau)-\overline{S}_m^{(1)}(\tau/2)]\},
\end{equation}
\end{widetext}
\newpage 
where we introduce the following shorthand notation involving quantities $\overline{C}_m^{(i)}(t)$,
$C_m^{(i)}(t)$, $\overline{S}_m^{(i)}(t)$ and $S_m^{(i)}(t)$
\newpage
\begin{align}
    C_{n}^{(1)}(t) &= \int_{0}^{t}e^{\gamma t'} \cos(\frac{4\pi n}{\tau} t') dt' \\
    \overline{C}_{n}^{(1)}(t) &= \int_{0}^{t}e^{-\gamma t'} \cos(\frac{4\pi n}{\tau} t') dt' \\
    S_{n}^{(1)}(t) &= \int_{0}^{t}e^{\gamma t'} \sin(\frac{4\pi n}{\tau} t') dt' \\
    \overline{S}_{n}^{(1)}(t) &= \int_{0}^{t}e^{-\gamma t'} \sin(\frac{4\pi n}{\tau} t') dt' \\
\end{align}
For the particular set of drivings from Eq. (\ref{force_trigonometric}),and considering $\omega_j=2 \pi/\tau$, Onsager coefficients reduce to the following expressions:
\begin{widetext}
\begin{align}
   L_{11} =\frac{mT\tau  \left[\gamma ^3 \tau ^3+4 \pi ^2 \gamma  \tau +16 \pi ^2 \coth \left(\frac{\gamma  \tau
   }{4}\right)\right]}{4 \left(\gamma ^2 \tau ^2+4 \pi ^2\right)^2},
\end{align}
%

\begin{align}
 L_{22}=\frac{mT\tau  \left[\gamma  \tau  \left(\left(e^{\frac{\gamma  \tau }{2}}-1\right) \left(\gamma ^2 \tau ^2+4
   \pi ^2\right)-4 \gamma  \tau  \left(e^{\frac{\gamma  \tau }{2}}+1\right) \sin ^2(\phi )\right)+16 \pi ^2
   \left(e^{\frac{\gamma  \tau }{2}}+1\right) \cos ^2(\phi )\right]}{4 \left(e^{\frac{\gamma  \tau
   }{2}}-1\right) \left(\gamma ^2 \tau ^2+4 \pi ^2\right)^2},
\end{align}
%

%
\begin{align}
    L_{12} = 
    -\frac{2 \pi mT \tau  \coth \left(\frac{\gamma  \tau }{4}\right) (\gamma  \tau  \sin (\phi )+2 \pi  \cos (\phi
   ))}{\left(\gamma ^2 \tau ^2+4 \pi ^2\right)^2},
\end{align}
and
%
\begin{align}
    L_{21} =\frac{2 \pi mT \tau  \coth \left(\frac{\gamma  \tau }{4}\right) (\gamma  \tau  \sin (\phi )-2 \pi  \cos (\phi
   ))}{\left(\gamma ^2 \tau ^2+4 \pi ^2\right)^2},
\end{align}
\end{widetext}
respectively.
%

\section{Onsager coefficients for power-law drivings}\label{appa}

For generic algebraic (power-law) drivings, Onsager coefficients
are listed below:
\begin{widetext}
\begin{equation}
    L_{11} = \frac{mT}{ \tau  } \int_{0}^{ \tau/2  } \Bigg[ 4^{\alpha } e^{-t} \left(\frac{t}{\tau }\right)^{\alpha } \left(\frac{(-\tau )^{-\alpha } \left(\Gamma \left(\alpha +1,-\frac{\tau }{2}\right)-\Gamma (\alpha
   +1)\right)}{e^{\tau }-1}+(-t)^{-\alpha } \left(\frac{t}{\tau }\right)^{\alpha } (\Gamma (\alpha +1,-t)-\alpha  \Gamma (\alpha ))\right) \Bigg] dt,
\end{equation}
\begin{equation}
   L_{12}= \frac{mT}{ \tau  } \int_{0}^{ \tau/2 } \Bigg[ (-1)^{\beta } e^{-t} 2^{\alpha +\beta -1} (-\tau )^{-\beta } \text{csch}\left(\frac{\tau }{2}\right) \left(\frac{t}{\tau }\right)^{\alpha } \left(\Gamma
   \left(\beta +1,-\frac{\tau }{2}\right)-\Gamma (\beta +1)\right) \Bigg] dt,
\end{equation}
%
%
\begin{equation}
   L_{21}= \frac{mT}{ \tau  } \int_{ \tau/2 }^{ \tau } \Bigg[ \frac{2^{\alpha } (-\tau )^{-\alpha } e^{\tau -t} \left(1-\frac{2 t}{\tau }\right)^{\beta } \left(\Gamma \left(\alpha +1,-\frac{\tau }{2}\right)-\Gamma (\alpha
   +1)\right)}{e^{\tau }-1} \Bigg] dt,
\end{equation}
and
\begin{equation}
    L_{22} = -\frac{mT}{ \tau (e^\tau-1)  } \int_{ \tau/2 }^{ \tau } \Bigg\{ e^{\frac{\tau }{2}-t} \left(1-\frac{2 t}{\tau }\right)^{\beta } \left[\left(\frac{2}{\tau}\right)^\beta \left(\Gamma (\beta +1)-\Gamma \left(\beta
   +1,-\frac{\tau }{2}\right)\right)+\left(e^{\tau }-1\right) \left(2-\frac{4 t}{\tau }\right)^{\beta } (\tau -2 t)^{-\beta } \left(\Gamma (\beta +1)-\Gamma
   \left(\beta +1,\frac{\tau }{2}-t\right)\right)\right] \Bigg\} dt,
\end{equation}
\end{widetext}
respectively, where $\Gamma(x)$ and $\Gamma(x,y)$ denote gamma and incomplete gamma functions, respectively.
\bibliography{refs}

\end{document}